\documentclass[twocolumn,showpacs,prl]{revtex4-1}

\usepackage{graphicx}
\usepackage{amsmath}\usepackage{amssymb}\usepackage{multirow}
\usepackage{float}

\begin{document}

\title{Majorana Fermions and Disclinations in Topological Crystalline Superconductors}

\author{Jeffrey C.Y. Teo}
\author{Taylor L. Hughes}
\affiliation{Department of Physics, University of Illinois at Urbana-Champaign, IL 61801, USA}

\begin{abstract}
We prove a topological criterion for the existence of zero-energy Majorana bound-state on a disclination, a rotation symmetry breaking point defect, in 4-fold symmetric topological crystalline superconductors (TCS). We first establish a complete topological classification of TCS using the Chern invariant and three integral rotation invariants. By analytically and numerically studying disclinations, we algebraically deduce a $\mathbb{Z}_2$-index that identifies the parity of  the number of Majorana zero-modes at a disclination. Surprisingly, we also find weakly-protected Majorana fermions bound at the corners of superconductors with trivial Chern and weak invariants.
\end{abstract}

\maketitle
Symmetry protected topological insulators and superconductors have theoretically, and experimentally, risen to prominence in the last half-decade\cite{KaneHasan}. Recent developments in this field have moved on from the study of discrete symmetries such as time-reversal and charge-conjugation\cite{SchnyderXX,QiHugesZhang,KitaevKtheory}, to translational and point group symmetries\cite{FuKane2,TeoFuKane,Fu,HugesProdanBernevig,Turner,SnTe,PbSnTe,PbSnTe2,FangGilbertBernevig,Chernconstraint}. While spatial symmetries are not preserved as generically as, say, time-reversal, they can still support robust topological states in clean crystalline systems. It is understood that the so-called \emph{strong} topological invariants, which are protected no-matter what spatial symmetries are broken, determine the appearance of disorder-insensitive gapless boundary states. Interestingly, it was found that \emph{weak} invariants, which require an additional translation symmetry, support boundary states\cite{Ringel,Mong}, and more surprisingly, robust bound-states on crystal dislocations\cite{dislocationTI,TISCdefect,Ran,Nagaosa,Zaanen}. A natural question is then to ask whether topological defects of the point-group rotational symmetry, \emph{i.e.} disclinations, can also bind low-energy states in topological phases protected by point-group symmetry. A related problem has been studied in a different context in graphene\cite{graphenegaugefield,graphenedisclination,grapheneTIdisclination}. In this Letter we address this question for 2d topological \emph{superconductors} with point-group symmetry. To determine the existence of zero-energy Majorana bound states (MBS)\cite{Ivanov,ReadGreen} on disclinations we derive an index theorem connecting the parity of the number of zero modes to the eigenvalues of rotation operators at rotationally invariant points in the Brillouin zone (BZ). Our derivation combines a unique algebraic approach with exact diagonalization numerics and can be generalized to prove generic index theorems. 

\begin{figure}[t]
	\centering
	\includegraphics[width=3.5in]{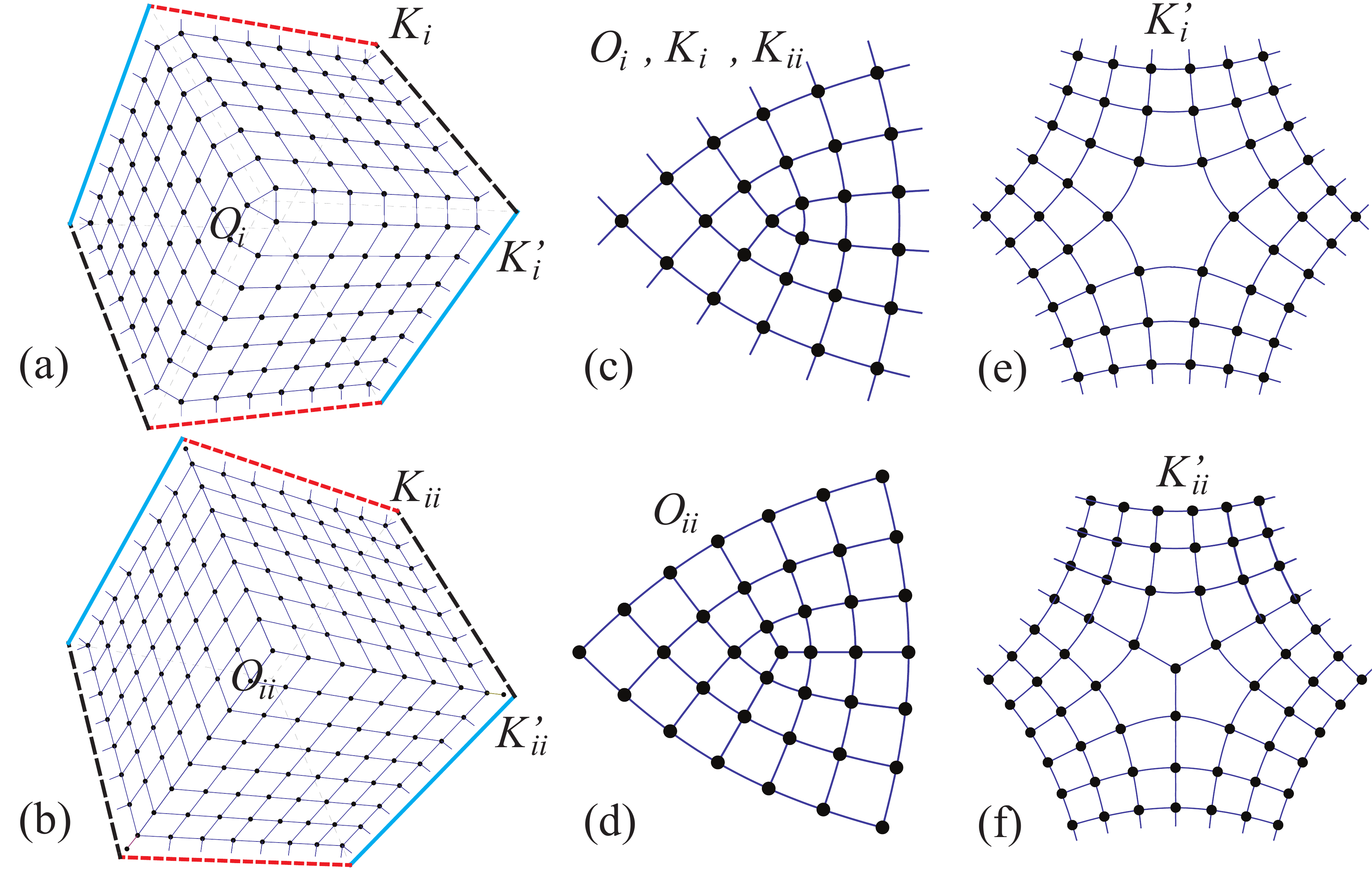}
 \caption{(a), (b)Lattice configurations with three disclinations. Each of the three faces is a $L\times L$ square lattice. Periodic boundary conditions are taken on the six edges of the cubes as indicated by matching colors/line styles. (c)-(f) Flattened, zoomed-in versions of the $-90^\circ$ (c,d) and $+180^\circ$ (e,f) disclinations at $O$, $K$ and $K'$. (c) {\em type-1} disclination centered at a triangular plaquette with an odd number of translations around its boundary. (d) {\em type-0} disclination centered at a trivalent vertex with an even number.}\label{disclination}
\end{figure}

Just as dislocations are local topological defects in the translational order of crystals, disclinations are topological defects in the discrete rotational order. In 2d, disclinations are point defects that can be constructed by a Volterra process\cite{Volterra} of removing or inserting material in certain angular sections with angles compatible with the crystalline symmetry (see Fig. \ref{disclination}). \emph{Dislocations} are characterized by their Burgers' vector \emph{i.e.} the translation element acquired when a particle encircles the defect; \emph{disclinations} are described by an element in the space group encoding the amount of rotation $\Omega$ \emph{and} translation ${\bf T}$ one picks up traveling around the point defect. The translation piece ${\bf T}$ is not unique and depends on the enclosing path,  however, for a $C_4$ symmetric lattice the evenness (type-0) or oddness (type-1) of the number of translations is unique. The $C_4$ symmetry thus yields a $\mathbb{Z}_2$-characteristic that distinguishes classical disclinations with the same Frank angle $\Omega.$ Examples are shown for $-90^\circ$ disclinations in Fig. \ref{disclination}c,d. The difference between the two primitive disclinations can be observed at the defect cores where $C_4$-symmetry is violated at either a triangular plaquette or a trivalent vertex. The $\mathbb{Z}_2$ characteristic arises from the fact that there are two inequivalent 4-fold rotation centers, vertex or plaquette, which can be extended to general $C_n$\cite{disclinationconjugacy}. 

We will only consider fully-gapped, translationally invariant superconductors in the mean-field limit which are described by Bogoliubov-de Gennes (BdG) Hamiltonians $H({\bf k})$, in Bloch form, with a particle hole constraint $\Xi H({\bf k})\Xi^\dagger=-H(-{\bf k})$, where $\Xi$ is a local, anti-unitary operator. A point group element $r$ is represented by a unitary operator $\hat{r}$ that commutes with the full Hamiltonian, and satisfies $\hat{r}H({\bf k})\hat{r}^\dagger=H(r{\bf k})$ for the Bloch Hamiltonian, and $\Xi\hat{r}\Xi^\dagger=\hat{r}$ for the particle-hole operator\cite{localFP}. For the duration of our work we will focus on the Abelian point-group $C_4$, generated by $\pi/2$-rotations, which is a symmetry commonly shared by all layered perovskites. We use the half-integer spin rotation such that $\hat{r}^4=(\hat{r}^2)^2=-1.$

$C_4$-symmetric superconductors in two dimensions is classified by (i) the Chern invariant \begin{equation}ch=\frac{i}{2\pi}\int_{BZ}d^2 k\;\epsilon^{ij}\mbox{Tr}(\partial_{k_i}\mathcal{A}_j)\in\mathbb{Z}\label{ch}\end{equation} for $(\mathcal{A}_{i})_{mn}({\bf k})=\langle u_m({\bf k})|\partial_{k_i}\vert u_n({\bf k})\rangle$ being the Berry connection of the negative-energy bands\cite{TKNN,Volovik,Kitaevanyon}, (ii) the eigenvalues of the rotation operator $\hat{r}$ for all the negative energy states at the 4-fold symmetric momenta $\Pi=\Gamma, M(=(\pi,\pi))$ (iii) the spectrum of the $C_2$ rotation $\hat{r}^2$ at \emph{one} of the \emph{equivalent} 2-fold symmetric momenta $X=(\pi,0)$ or $X'=(0,\pi).$ We label the bands at the symmetry points by their rotation eigenvalues following Ref. \onlinecite{PGnotation}. At $\Pi=\Gamma,M$, a band with $\hat{r}=e^{-i\pi/4},e^{i\pi/4},e^{3i\pi/4},e^{-3i\pi/4}$ is labelled by $\Pi_5,\Pi_6,\Pi_7,\Pi_8$ respectively; while at $X$, a band with $\hat{r}^2=i,-i$ is labelled by $X_3,X_4$ respectively. Let $\#\Pi_i,\#X_i$ be the number of appearances of $\Pi_i,X_i$ within the negative-energy states. Only differences of the eigenvalues between symmetry points carry topological information. Our convention uses the eigenvalues relative to the values at the $\Gamma$-point, and we define:
\begin{eqnarray}n_3&=&\#X_3-\#\Gamma_6-\#\Gamma_8\\n_4&=&\#X_4-\#\Gamma_5-\#\Gamma_7\\n_i&=&\#M_i-\#\Gamma_i,\quad\mbox{for $i=5,6,7,8$}.\end{eqnarray} These are easy to understand as $\Gamma_6,\Gamma_8$ ($\Gamma_5,\Gamma_7)$ both square to $+i (-i)$ and thus $n_3, n_4$ determine the difference in the $C_2$ eigenvalues at $X$ and $\Gamma$ while $n_{5,6,7,8}$ determine the $C_4$ eigenvalue differences between $M$ and $\Gamma.$ Including $ch$ this gives seven topological numbers, but there are constraints which imply the topology is determined by fewer quantities. 
These integers obey $n_3+n_4=n_5+n_6+n_7+n_8=0$ (from the constant number of negative-energy bands throughout the BZ) and $n_5+n_6=n_7+n_8=0$ (from the particle-hole constraint ensuring the rotation spectrum at symmetry points of the unoccupied bands is the complex conjugate of that of the occupied ones). Following the work of Ref. \onlinecite{Chernconstraint} one can show that 
 \begin{equation}ch+n_6+2n_4+3n_7\equiv0\quad\mbox{mod $4$}.\label{chconstraint}\end{equation} Hence $C_4$-symmetry topological crystalline superconductors are completely classified by four integral invariants $\chi_i\equiv (ch,n_4,n_6,n_7)$ that satisfy \eqref{chconstraint}. Moreover the {\em weak} $\mathbb{Z}_2$-topological invariant is determined by the inversion eigenvalues at $M$ and $X$: \begin{eqnarray}{\textbf{G}}_\nu=\nu({\bf G}_1+{\bf G}_2),\quad\nu=(n_4+n_6-n_7)\;\mbox{mod $2$}\label{w2}\end{eqnarray} where ${\bf G}_1,{\bf G}_2$ are the reciprocal lattice vectors.\cite{2ndweak}

\begin{figure}[t]
 \centering
	\includegraphics[width=2.7in]{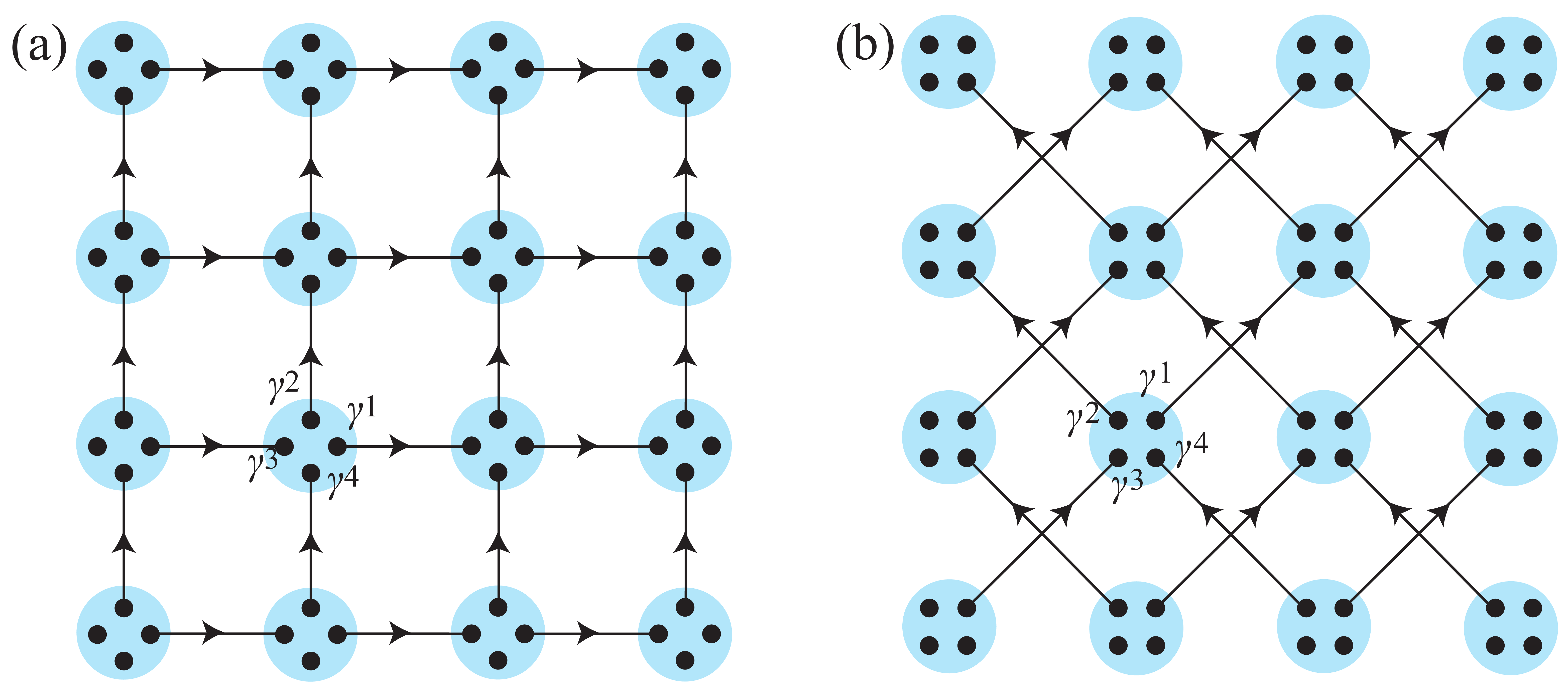}
 \caption{$C_4$-symmetric tight binding models with four Majorana fermions (black dots) at each site. (a)  $H_b$ (b) $H_c$}\label{atomicmodel}
\end{figure} 

The essential ingredient of our index-theorem proof is a collection of 2d $C_4$ symmetric superconductor models that ``generate" all the topological classes characterized by the $\chi_i.$ Since $\chi_i$ is a four component vector we need four Hamiltonians which have linearly independent $\chi_i.$ Combining the Hamiltonians via direct sum combines the vectors with a vector sum, so any topological class $\chi_i$ can be produced. We choose two generators to be spinless, chiral $p_x+ip_y$ superconductors on a square lattice: 
\begin{eqnarray}H_a&=&\Delta(\sin k_x\tau_x+\sin k_y\tau_y)\nonumber\\&&+[u_1(\cos k_x+\cos k_y)+2u_2\cos k_x\cos k_y]\tau_z\label{Hc}\end{eqnarray} where $\tau_a$ acts on Nambu-space, $\Delta$ is the $p_x+ip_y$ pairing and the first/second neighbor hoppings $u_1, u_2$ are kinetic energy terms that gap out the nodes of the pairing term at the points $\Gamma,X,X',M.$ The particle-hole and rotation operators are given by $\Xi_a=\tau_xK$ and $\hat{r}_a=(\openone_2+i\tau_z)/\sqrt{2}$ where $K$ is complex conjugation. The invariants $\chi_i$ depend on $u_1$ and $u_2$ and are summarized in Table \ref{tb2}. Flipping the signs of both $u_1$ and $u_2$ inverts $\chi_i\to -\chi_i.$ 
\begin{table}[t]
\centering
\begin{ruledtabular}
\begin{tabular}{c|c|cccc}
$H_a$  & hopping strength & $ch$ & $n_4$ & $n_6$ & $n_7$\\
\hline
$H_a^{(1;0)}$ & $u_1>u_2>0$ & 1 & $-1$ & 1 & 0\\
$H_a^{(1;1)}$ & $-u_1>u_2>0$ & 1 & 0 & $-1$ & 0\\
$H_a^{(2;1)}$ & $u_2>|u_1|$ & 2 & $-1$ & 0 & 0\\
\end{tabular}
\end{ruledtabular}
\caption{Chern and rotation invariants for the chiral $p_x+ip_y$ superconductor \eqref{Hc}. The Hamiltonians are superscript labeled by their Chern and weak invariants $(ch;\nu)$.}\label{tb2}
\begin{ruledtabular}
\begin{tabular}{c|cccc}
TB model  & $ch$ & $n_4$ & $n_6$ & $n_7$\\
\hline
$H_b$ & 0 & 1 & $-1$ & 1\\
$H_c$ & 0 & 2 & 0 & 0\\
\end{tabular}
\end{ruledtabular}
\caption{Chern and rotation invariants of models in Fig. \ref{atomicmodel}.}\label{tb1}
\end{table}

The other independent generators are 2d generalizations of Kitaev's p-wave wire\cite{Kitaevchain}. Fig. \ref{atomicmodel} depicts two tight-binding limits of $C_4$ symmetric Majorana fermion models with four fermions per site and arrows which represent the Majorana ordering convention. The two Hamiltonians are $\hat{H}_b=it\sum_{\bf x}\left(\gamma^1_{\bf x}\gamma^3_{{\bf x}+e_1}+\gamma^2_{\bf x}\gamma^4_{{\bf x}+e_2}\right)$ and $\hat{H}_c=it\sum_{\bf x}\left(\gamma^1_{\bf x}\gamma^3_{{\bf x}+e_1+e_2}+\gamma^2_{\bf x}\gamma^4_{{\bf x}-e_1+e_2}\right)$, where $\gamma^{i}_{\bf x}$'s are Majorana operators with $\gamma_{\bf x}^{i\dagger}=\gamma_{\bf x}^{i}$ and $\{\gamma^i_{\bf x},\gamma^j_{\bf y}\}=2\delta^{ij}\delta_{\bf xy}$. The $C_4$ rotation operator $\hat{r}_{bc}=\prod_{\bf x}\exp(-\frac{\pi}{4}\gamma^1_{\bf x}\gamma^2_{r{\bf x}})\exp(-\frac{\pi}{4}\gamma^2_{\bf x}\gamma^3_{r{\bf x}})\exp(-\frac{\pi}{4}\gamma^3_{\bf x}\gamma^4_{r{\bf x}})$ gives $\hat{r}_{bc}(\gamma^1_{\bf x},\gamma^2_{\bf x},\gamma^3_{\bf x},\gamma^4_{\bf x})\hat{r}_{bc}^{\dagger}=(\gamma^2_{r{\bf x}},\gamma^3_{r{\bf x}},\gamma^4_{r{\bf x}},-\gamma^4_{r{\bf x}})$, where $r$ is the $C_4$ rotation in real space. If we transform to complex fermions $c=(\gamma^1+i\gamma^3)/2$ and $d=(\gamma^2+i\gamma^4)/2$  then the corresponding BdG Hamiltonians take the block diagonal forms
\begin{eqnarray}
 H_b({\bf k})&=&t(\cos k_x\tau_z+\sin k_x\tau_y)\oplus t(\cos k_y\tau_z+\sin k_y\tau_y)\nonumber\\
 H_c({\bf k})&=&t(\cos(k_x+k_y)\tau_z+\sin(k_x+k_y)\tau_y)\nonumber\\
 &\oplus& t(\cos(k_x-k_y)\tau_z+\sin(k_y-k_x)\tau_y)
 \end{eqnarray}
in the basis $\vec\xi_{\bf k}=(c_{-{\bf k}},c^\dagger_{\bf k},d_{-{\bf k}},d^\dagger_{\bf k})^T$ and where $\tau_i$'s act on the Nambu degree of freedom. The particle-hole and rotation operators are $\Xi_{bc}=(\openone_2\otimes\tau_x)K$ and $\hat{r}_{bc}=\sigma_+\otimes\openone_2-i\sigma_-\otimes\tau_z$ where $\sigma_{\pm}=1/2(\sigma_x\pm i \sigma_y)$   acts on the $(c,d)$ space.
The lattice structure of $\hat{H}_c$ is two copies of $\hat{H}_b$ displaced by half a lattice spacing, however, as a $C_4$-symmetric electronic structure,  $H_c({\bf k})$ is not two copies of $H_b({\bf k})$ because of the different rotation centers. This is evident in the respective $\chi_i$ shown in Table \ref{tb1}. Together with the two tight binding models  $H_a^{(1;0)}$ and $H_a^{(1;1)}$, $H_b$ and $H_c$  generate all possible combinations of $\chi_i=(ch,n_4,n_6,n_7)$ allowed by \eqref{chconstraint}.  Thus, every $C_4$-symmetric superconductor must have the identical \emph{topological} properties to direct sums of the two $H_a$'s, $H_b,$ and $H_c.$ In particular the zero-modes at disclinations of \emph{any} $C_4$-symmetric superconductor can be determined from these four models.

\begin{figure}[t]
 \centering
	\includegraphics[width=3.5in]{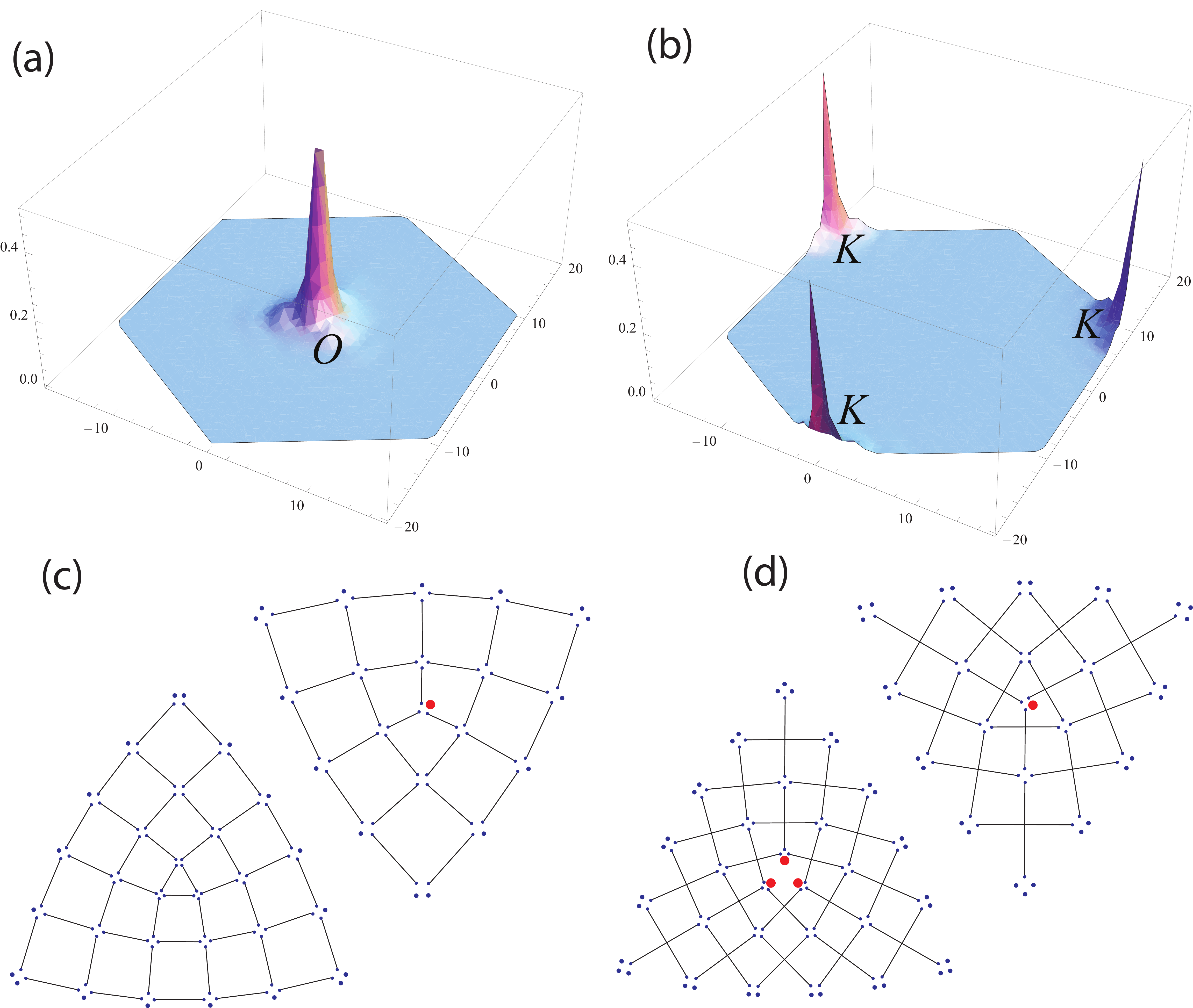}
 \caption{(a,b) Exponentially localized probability amplitudes of the Majorana zero mode at disclinations $O$ and $K$, plotted on a torus geometry where parallel sides on the hexagonal domain are identified. Compare with 3d lattice in Fig. \ref{disclination}. (c) Tight binding model $H_b$ with (left) type-1 and (right) type-0 disclinations (d) $H_c$ with (left) type-1 (right) type-0 disclinations. Thick red dots in disclination cores are unpaired Majorana boundstates. }\label{plots}
\end{figure}

After constructing the Hamiltonian generators, we must determine the properties of disclinations for each generator.
A disclination configuration is specified by the pair $(\Omega,{\bf T})$ where $\Omega$ is the Frank angle and ${\bf T}$ is the translation. 
The MBS of the chiral superconductors $H_a$ (with pairing and hopping parameters $2u_2/\Delta=\pm u_1/\Delta=1$) are studied numerically using periodic lattice models with three disclinations. This is achieved by taking three adjacent faces of a cube and gluing the parallel sides (see Fig. \ref{disclination}a,b). Two $-90^\circ$ disclinations are located at the points $O$ and $K$ of positive curvature and a $+180^\circ$ disclination is located at the point $K'$ of negative curvature. We use two lattice configurations that differ in the choice of  a type-0 or type-1 disclination at $O$ (see Fig.\ref{disclination}c,d). The  superconducting phase is chosen so that it smoothly winds around the defects, but there are two inequivalent ways of specifying the defect lattice due to the double covering of the rotation group. The smooth winding around the disclination involves either the rotation $\hat{r}(s)=e^{is\Omega \tau_z/2}$ \emph{or} $\hat{r}'(s)=e^{is(2\pi+\Omega)\tau_z/2}$, parameterized by $s\in[0,1].$  We choose $\hat{r}(s)$ and thus the phase smoothly winds by $\pi/2$ around $O$, $K$ and $-\pi$ around $K'.$ Note, if we choose $\hat{r}'(s)$ then the corresponding 4-fold operator would be $-\hat{r}$ instead and the invariants are flipped, $n_6\leftrightarrow-n_7$; the index is then calculated the same way after this change. We find that only $H_{a}^{(1;1)}$ supports an odd number of MBS and even then only for type-1 disclinations. The results are summarized in Table \ref{tb3} and we show the zero mode wave functions at the disclinations at $O$ and $K$  in Fig. \ref{plots}a,b.

For $H_b, H_c$ we use the fact that the parity of the number of MBS at the defect is insensitive to all perturbations that do not violate the energy gap or rotation symmetry away from the point defect since there is no low-energy channel for a single MBS to escape or enter the disclination core. This implies that, just as for the boundary modes of the topological p-wave wire\cite{Kitaevchain}, we can determine the parity of the zero-mode bound states pictorially in the tightly-bound limit.
The MBSs of $H_b$ and $H_c$ at  type-0 and type-1 $-90^\circ$ disclinations are shown in Fig. \ref{plots}c,d, where the MBSs are simply un-bonded Majorana fermions represented by thick red dots. 
We find that $H_b$ has a zero-mode for type-0, and $H_c$ has zero modes for both types. This is summarized in Table \ref{tb3}.

\begin{table}[t]
\centering
\begin{ruledtabular}
\begin{tabular}{c|cccc}
$-90^\circ$ disclination  & $H_a^{(1;0)}$ & $H_a^{(1;1)}$ & $H_b$ & $H_c$\\
\hline
type-0 & 0 & 0 & 1 & 1\\
type-1 & 0 & 1 & 0 & 1\\
\end{tabular}
\end{ruledtabular}
\caption{Parity of the number of zero modes at a $-90^\circ$ disclination for the chiral superconductors $H_a^{(1;0)}, H_a^{(1;1)}$ in \eqref{Hc} with smooth rotation $\hat{r}(s)=e^{is\Omega\tau_z/2}$ and the tight binding models $H_b, H_c$ in Fig. \ref{atomicmodel}.}\label{tb3}
\end{table}

We are now in a position to determine the index theorem since any $C_4$ symmetric BdG hamiltonian can be smoothly deformed into a unique composition \begin{equation}[H]\simeq m_1[H_{a}^{(1;0)}]\oplus m_2[H_{a}^{(1;1)}]\oplus m_3[H_b]\oplus m_4[H_c]\end{equation} up to topologically trivial bands far from Fermi-level. The $m_i$ are integers, and $m_i<0$ means a direct sum with the negative and positive energy states switched. To finish the derivation of the topological index we need to carry out some simple algebraic manipulations. First, we see that since $H_c$ only has $n_4=2$ non-zero and has zero-modes for both disclination types we determine that the index gets a contribution of $1/2(n_4)\;\mbox{mod 2}.$ Next we can take $2H_{a}^{(1;0)}\oplus 2H_{a}^{(1;1)}\oplus H_c$ which has $\chi_i=(4,0,0,0)$ and bound-states for both types. This implies the index recieves a contribution of $1/4(ch)\;\mbox{mod 2}.$ Using these two pieces we can go back to $H_{a}^{(1;0)},$ which does not have any zero-modes, and determine the equation $[1/2(n_4)+1/4(ch)+k(n_6)]\;\mbox{mod 2}=0$ which upon substitution gives $k=1/4.$ To determine the contribution of $n_7$ we consider $H_{a}^{(1;0)}\oplus H_{a}^{(1;1)}\oplus H_b$ with $\chi_i=(2,0,-1,1).$ This model has bound states on both types of disclinations so we use $[1/2(n_4)+(1/4)(ch)+(1/4)n_6+jn_7]\;\mbox{mod 2}=1$ to find $j=3/4.$ So far we were careful to choose all the Hamiltonian combinations above to have a vanishing weak invariant as it also contributes to the index. We can see this by taking $H_{a}^{(1;1)}$ which yields $1/4[ch+n_6+2n_4+3n_7]\;\mbox{mod 2}=0,$ yet has a bound state on type-1 disclinations. This bound state, however arises from a different mechanism, and comes from the interplay of the non-zero weak invariant  and the oddness of the translation ${\bf T}$ around a type-1 disclination. Thus, we have determined the existence conditions for an odd number of MBSs at a $-90^\circ$ disclination. Combining disclinations gives $(\Omega_1,{\bf{T}}_1)+(\Omega_2,{\bf{T}}_2)=(\Omega_1+\Omega_2, {\bf{T}}_1 + r(\Omega_1){\bf{T}}_{2})$ which implies that for generic $C_4$ disclinations with Frank angle $\Omega$ the topological index is: 
\begin{equation}\Theta\equiv\left[\frac{1}{2\pi}{\bf T}\cdot{\bf G}_\nu+\frac{\Omega}{2\pi}\left(ch+n_6+2n_4+3n_7\right)\right]\;\;\mbox{mod 2}\label{index1}\end{equation} where ${\bf G}_\nu$ is the weak $\mathbb{Z}_2$ invariant. As mentioned, the first term resembles the topological index for MBSs at a dislocation, where the Burgers' vector ${\bf B}$ is replaced by ${\bf T}$\cite{dislocationTI,TISCdefect,Ran}. It vanishes for type-0 disclinations and equals $n_4+n_6+n_7$ (mod 2) for type-1's. The second term of \eqref{index1} is an integer because of the constraint \eqref{chconstraint} on the Chern invariant and can distinguish Chern numbers which are even or odd multiples of four. 

For a more physical understanding we can refer to the outer boundaries of the regions surrounding the disclinations in Fig. \ref{plots}c,d. For Fig. \ref{plots}c we see the boundary links carry one unbound Majorana and the corners contain two whereas in Fig. \ref{plots}d the links carry two and the corners carry three. Since there is only one disclination, the parity of MBSs on the boundary will match the parity in the disclination core. We can clearly see that the two terms that comprise $\Theta$ represent the \emph{edge} and \emph{corner} contributions to the index respectively. ${\bf T}\cdot{\bf G}_\nu/2\pi$ counts the number of MBSs (mod 2) on an edge with length ${\bf T}$, while $(ch+n_6+2n_4+3n_7)/4$ counts the number of MBSs at a $90^\circ$ corner. The index \eqref{index1} therefore not only gives information about the disclination core, but also the defect-free system boundary. In particular, even a system such as $H_c$, with vanishing Chern \emph{and} weak $\mathbb{Z}_2$ invariants (which thus does not carry topologically protected edge modes) binds Majorana fermions at corners since $\Theta\equiv1.$ This implies the existence of MBSs in the form of corner states even in a defect and vortex free system.
\begin{figure}[t]
 \centering
	\includegraphics[width=3.0in]{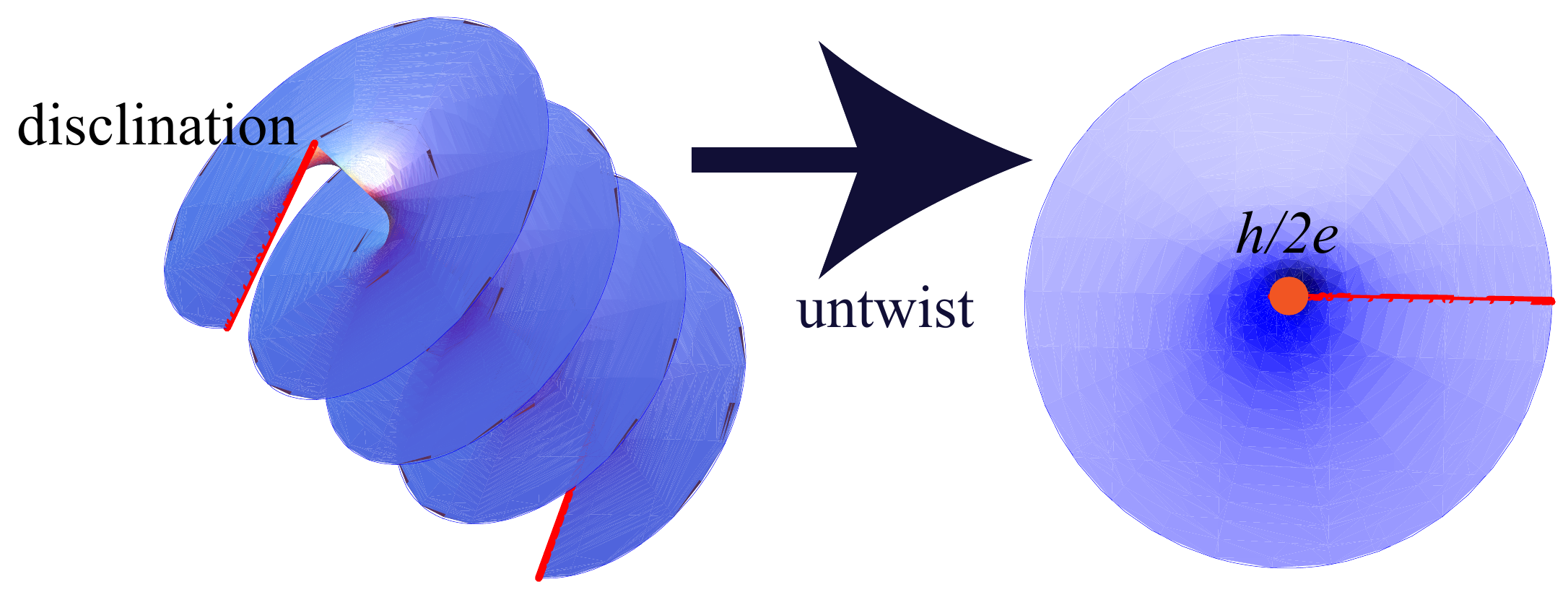}
 \caption{Untwisting a $-90^\circ$ disclination of the four layer $p_x+ip_y$ continuum model $H_4$ into a single layer $p_x+ip_y$ model with a quantum vortex.}\label{helix}
\end{figure}

$\Theta$ can be illustrated in a continuum system on a disk geometry with a disclination at the origin. Take four copies of a spinless, continuum $p_x+ip_y$ superconductor with each copy rotated by $\pi/2$ relative to the previous: $H_4=h_{0}\oplus h_{\pi/2}\oplus h_{\pi}\oplus h_{3\pi/2}$ for $h_\phi=e^{i\phi\tau_z/2}h_0e^{-i\phi\tau_z/2}$, $h_0({\bf k})=\vert\Delta\vert k_x\tau_z+\vert\Delta\vert k_y\tau_y+(m-\varepsilon k^2)\tau_z.$ $H_4$ has the discrete 4-fold rotation symmetry \begin{equation}\hat{r}_4=\left(\begin{array}{*{20}c}0&\openone_2&0&0\\0&0&\openone_2&0\\0&0&0&\openone_2\\-\openone_2&0&0&0\end{array}\right).\end{equation} A corresponding $-90^\circ$ disclination is represented by the helix in Fig. \ref{helix}, where the top and bottom layers are glued along the branch cut (red line) with anti-periodic boundary conditions since $\hat{r}_{4}^4=-1.$ The disclination helix can be untwisted to form a single copy of a $p_x+ip_y$ model with a $\pi$-flux \emph{vortex} replacing the disclination at the origin that binds a MBS\cite{ReadGreen}. This is consistent with the index theorem \eqref{index1} since $ch=4$, $n_4={\bf T}=0$ for continuum models, and also $n_6=n_7=0$ since momentum space can be compactified into a sphere $\mathbb{S}^2=\mathbb{R}^2\cup\{\infty\}$ where the rotation spectra at the fixed points $k=0$ and $\infty$ are identical.

We have shown that topological crystalline superconductors in two dimensions with $C_4$ rotation symmetry are classified by four integers $\chi_i=(ch,n_4,n_6,n_7).$ The appearance of zero-energy MBSs at disclinations is determined by a $\mathbb{Z}_2$ topological index $\Theta.$ Although the index theorem relies on $C_4$ symmetry, a MBS is robust against \emph{any} rotation breaking perturbation that does not close the bulk energy gap around the defect. In fact, the disclination itself breaks rotation symmetry. The MBS can be interpreted as a consequence of the topological nature of the BdG hamiltonian with a defect $H({\bf k},s)$\cite{TISCdefect} which does not assume any rotation symmetry (here $s$ adiabatically parametrizes a loop around the point defect). The MBSs at the corners of a sample, however, are sensitive to $C_4$ breaking perturbations, and can escape through accidental low-energy edge channels. If the edge reconstruction is weak enough, MBSs at corners could provide the possibility for indirect observation of the elusive excitation through transport experiments.

We thank Bryan Chen and Chen Fang for insightful discussions. JCYT was supported by the Simons Fellowship and TLH was supported by ONR award  N0014-12-1-0935.


\begin{thebibliography}{10}
\bibitem{disclinationconjugacy} For $C_2$ and $C_3$ there is a $\mathbb{Z}_2\times\mathbb{Z}_2$ and $\mathbb{Z}_3$ classification of disclinations with the same Frank angle respectively. $C_6$ only has one type.
\bibitem{localFP} In a lattice superconducting model, rotation is chosen to center at a unit cell that is compatible with local fermion parity (c.f. inversion center of the Kitaev superconducting chain\cite{Kitaevchain}).
\end{thebibliography}
\end{document}